\documentclass[12pt]{article}
\usepackage{amsmath, amssymb}
\usepackage{graphicx}
\usepackage{cite}
\usepackage{geometry}
\usepackage{authblk}
\usepackage{hyperref}
\title{New Magnetic Temperature Non-Contact Sensor}

\author[1]{Eddy Divin Kenvo Songwa}
\author[1]{Dima Cheskis\thanks{dimach@ariel.ac.il}}
\affil[1]{Department of Physics, Ariel University, Ariel 407000, Israel}

\date{}

\begin{document}

\maketitle

\begin{abstract}
Non-contact temperature sensors are widely used, often utilizing infrared light for temperature measurement. However, specific applications demand non-contact detection, particularly within closed containers containing fluids or gases, where optical methods are unsuitable. Our approach is designed precisely for this purpose. We conducted measurements, introduced a prototype of our detector, and confirmed its compatibility with nonmagnetic containers.
\end{abstract}


\section{Introduction}
Temperature measurement methods can be categorized into contact and non-contact. Contact methods include resistivity~\cite{Kuzubasoglu}, thermocouples~\cite{Turkiewicz}, and liquid thermometers, while non-contact methods involve using infrared radiation to measure temperature.
Non-contact methods include pyrometers  \cite{Araujo}, bolometers  \cite{Moisello}, and other infrared detectors, with surface acoustic wave detectors  \cite{Zhou} being a recent addition. For instance, in many cases, infrared body temperature sensors  \cite{Zhao} have replaced standard mercury thermometers  \cite{Haller}. In industrial applications, infrared thermometers are used in monitoring, equipment maintenance, electrical applications, and building control \cite{Usamentiaga,Balakrishnan,Zhang,Kylili,Huan}. In processes that involve liquid flow within pipes, it's crucial to use non-contact methods because they need to avoid invasive drilling into pipes, which could potentially lead to leaks. Non-contact infrared or surface acoustic thermal detectors are ineffective when it's challenging to precisely determine the temperature of the liquid within pipes, even if they are exposed to infrared radiation or acoustic waves emitted by the liquid.

\section{Sensor Design}
Our proposal involves the use of two distinct parts. An internal part consists of a bimetal strip that changes shape and becomes clamped due to temperature changes \cite{Maurelli}. Figure 1 demonstrates how this strip works. Two metals with differing thermal conductivities undergo distinct expansion when joined together—changes in temperature cause the combined strip to bend. By clamping one end and placing the other bimetal strip end inside a fluid, it is possible to indirectly measure the temperature inside the volume because the clamping of the bimetal strip is connected to a temperature change.
In the outside part of the sensor, we use a permanent magnet and a Hall sensor \cite{Castro} connected to a battery \cite{Sharon}. The magnet magnetizes a bimetal strip, and the Hall sensor detects changes in the combined magnetic field of the strip and magnet. Adjusting the distance between the strip and sensor changes the voltage amplitude in the Hall sensor, allowing us to determine the temperature change inside the container based on the measured magnetic field change outside. 

\begin{figure}[ht]
\centering
\includegraphics[width=6 cm, height=7cm]{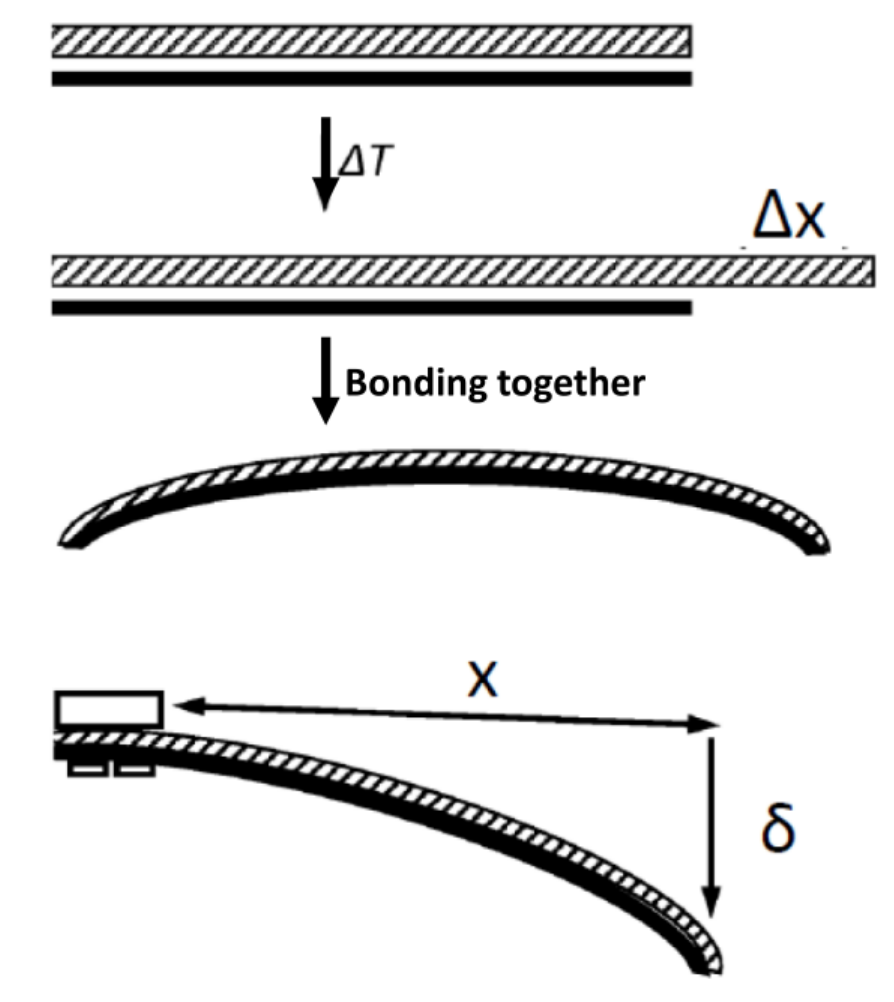}
\caption{The working principle of the bimetal strip.}
\label{fig:working_principle_bimetal_strip}
\end{figure}

A high signal-to-noise ratio (SNR) is necessary for accurate temperature measurement in the Hall sensor. To ensure the measuring system is independent of the power grid and can operate for an extended period, we utilize a battery-powered, low-constant control current Hall sensor, as demonstrated in Figure 2.

A non-contact temperature sensor finds valuable utility in water boilers \cite{Cheskis}. Temperature readings must be taken from various points within the boiler to regulate hot water heating. Installing sensors inside the boiler can be both challenging and expensive. Our sensor offers a prompt solution to this issue. As shown in Figure 3, a bimetallic strip is inserted into a boiler. A permanent magnet and a Hall sensor are positioned externally to the boiler. One end of the bimetallic strip is fixed to the inner surface of the container. When the water temperature rises, the bimetallic strip undergoes bending, causing the separation between the Hall sensor and the unattached end of the strip to increase. Initially, the permanent magnet magnetized this free end and contributed to the total magnetic field.

\begin{figure}[ht]
\centering
\includegraphics[width=\linewidth,height=5cm]{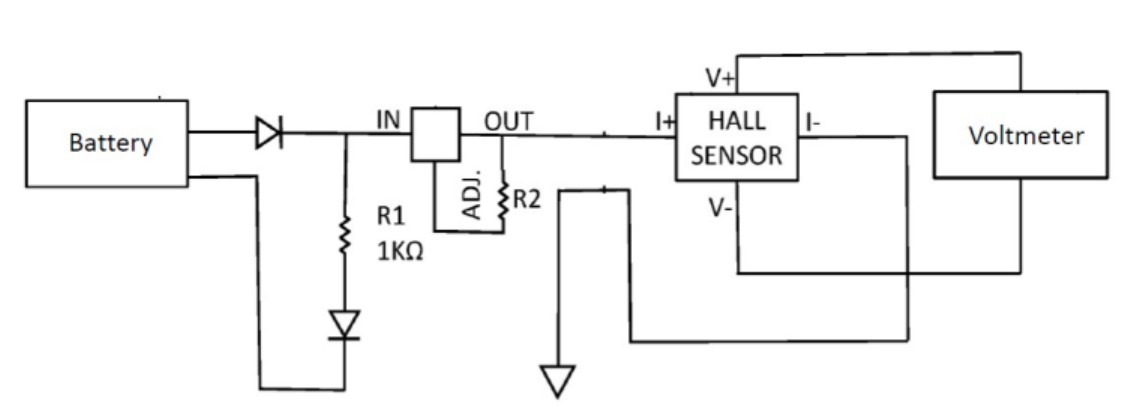}
\caption{The Hall system.}
\end{figure}

  The widening gap due to the temperature increase alters the overall magnetic signal detected by the Hall sensor. In the first stage, we plan the system, which checks the linearity of the system's bimetal layer placed near the heater element and our Hall sensor. We measured by heating a bimetallic strip to a high temperature and then gradually cooling it down.

\begin{figure}[ht]
\centering
\includegraphics[width=\linewidth,height=5.98cm]{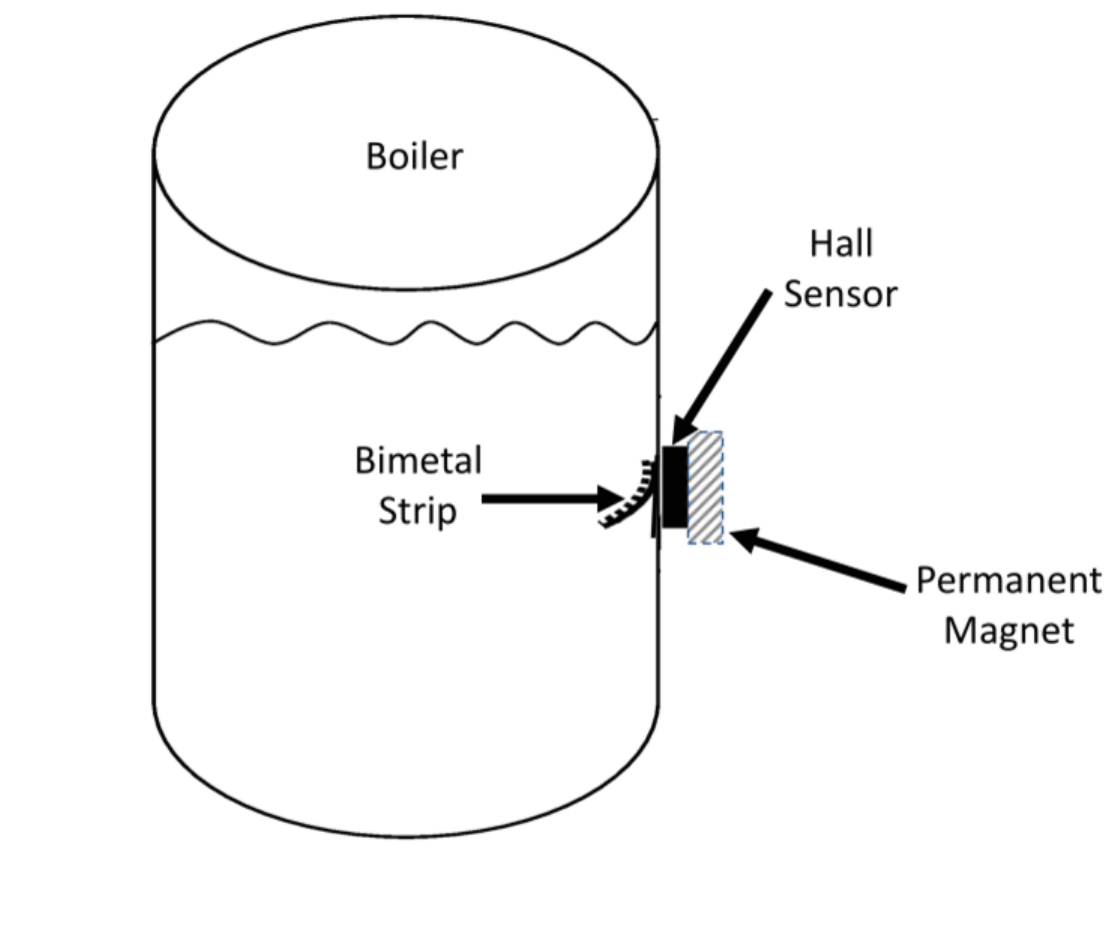}
\caption{The schematic description of water boilers temperature measurements.}
\end{figure}

\vspace{1\baselineskip}

\section{Experimental Setup}

In Figure 4, the left side illustrates our experimental plan, the top-right side shows our setup, and the bottom-right part displays the electrically powered heater.

\begin{figure}[ht]
\centering
\includegraphics[width=\linewidth,height=7.29cm]{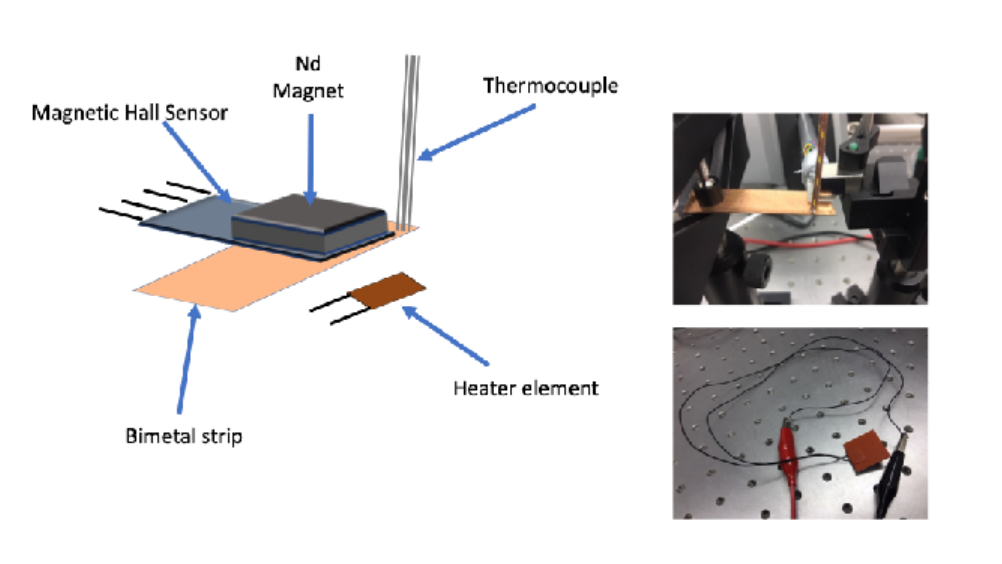}
\caption{The experimental system for non-contact temperature measurement.}
\end{figure}

 We aimed to replicate the previously described detector's functioning without using liquid and boiler walls. This initial step aimed to assess the detector's response linearity. To measure the temperature, we attached a thermocouple to the bimetallic strip, and the heater was affixed to the lower part of the strip. We removed the heater once the bimetallic strip curved significantly and reached 40 degrees Celsius. The measurement commenced at this point, with the bimetallic strip gradually cooling down while a thermocouple on the strip continuously monitored its temperature.

\section{Results and Discussion}

We perform measurements that simultaneously measure the change in temperature and the evolution of the magnetic signal. Results are shown in Figure 5. On the left side, we see a decrease in temperature as a function of time, which means that the bimetal strip cools down. At the same time, the Hall voltage is changing, which we see on the right side of Figure 5.

\vspace{1\baselineskip}
\begin{figure}[ht]
\centering
\includegraphics[width=\linewidth,height=6.8cm]{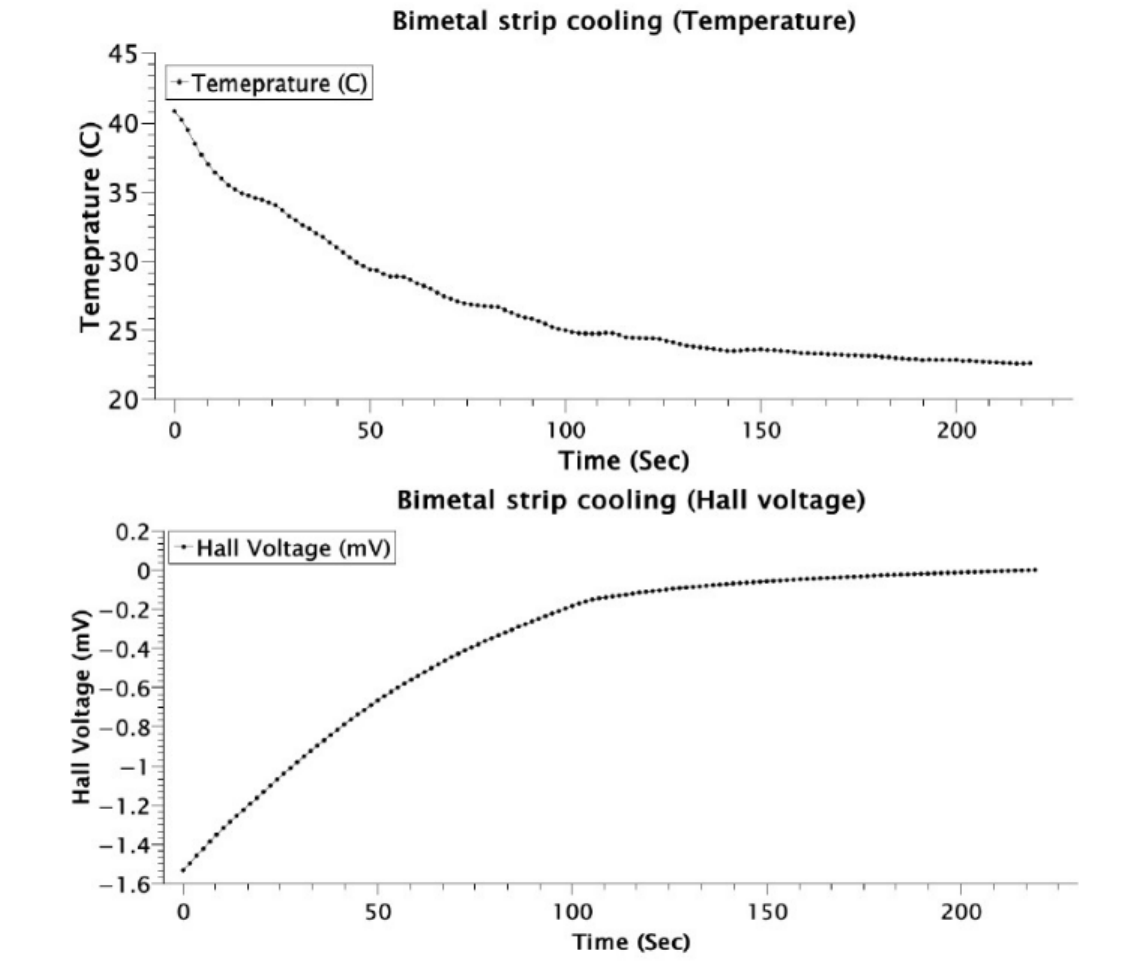}
\caption{The simultaneous change of magnetic signal and temperature of the bimetal strip during the cooldown process.}
\end{figure}

 These graphs look very similar. Because we measured these graphs simultaneously, we will check the Hall voltage dependence on the temperature. This dependence is shown in Figure 6.

On the left side of Figure 5, we observe a temperature decrease over time, indicating the cooling of the bimetal strip. On the right side of Figure 5, we show an increase in the Hall voltage depending on time. These two graphs exhibit striking similarities because we recorded them simultaneously. Consequently, we will examine the relationship between the Hall voltage and temperature variation, as illustrated in Figure 6.

\vspace{1\baselineskip}
\begin{figure}[ht]
\centering
\includegraphics[width=\linewidth,height=6.17cm]{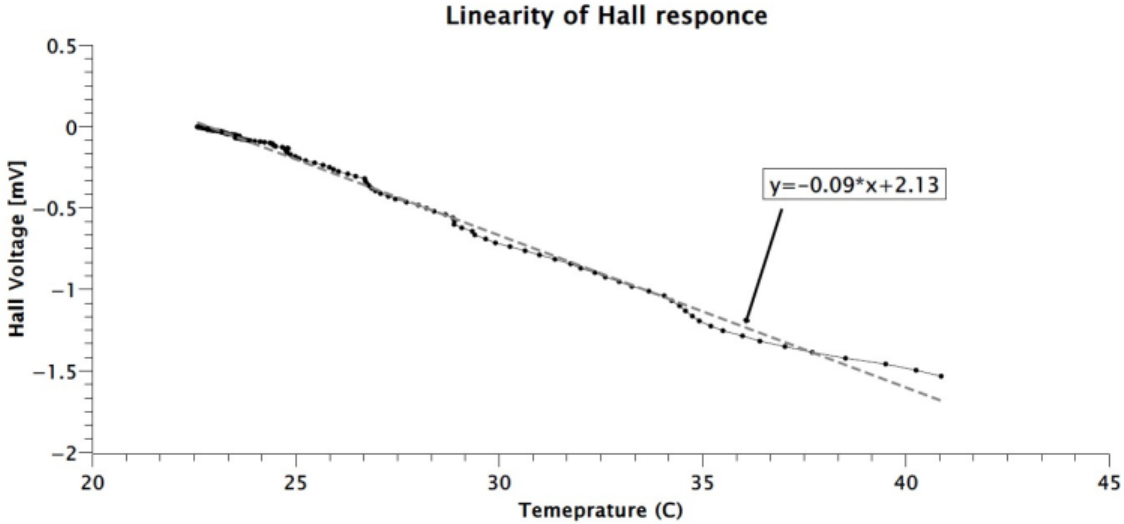}
\caption{The linear dependence between the magnetic sensor (Hall) voltage and temperature during the bimetal strip cooldown process.}
\label{fig:linear_dependence_magnetic_sensor_hall}
\end{figure}


As evident from the data, we acknowledged a linear relationship between temperature and Hall voltage. Despite minor fluctuations among data points that might not perfectly align on a straight line, a straightforward linear regression and correlation between the points are apparent when we perform a linear fit. These findings indicate the viability of using temperature sensors of this nature, even though alternative solutions may introduce complexities.

The next step is to check the response of the boiler wall made from steel and aluminum, as well as the response to temperature and magnetic field. The results are shown in Figures 7 and 8.

In Figure 7, we examine the response of steel. We observe that the relationship between external and internal temperatures is not highly linear, but linearity improves with temperature. Furthermore, the magnetic response in steel is relatively poor, likely due to interruptions in the ferromagnetic response lines within the magnetic field.

\begin{figure}[ht]
\centering
\includegraphics[width=\linewidth,height=5.86cm]{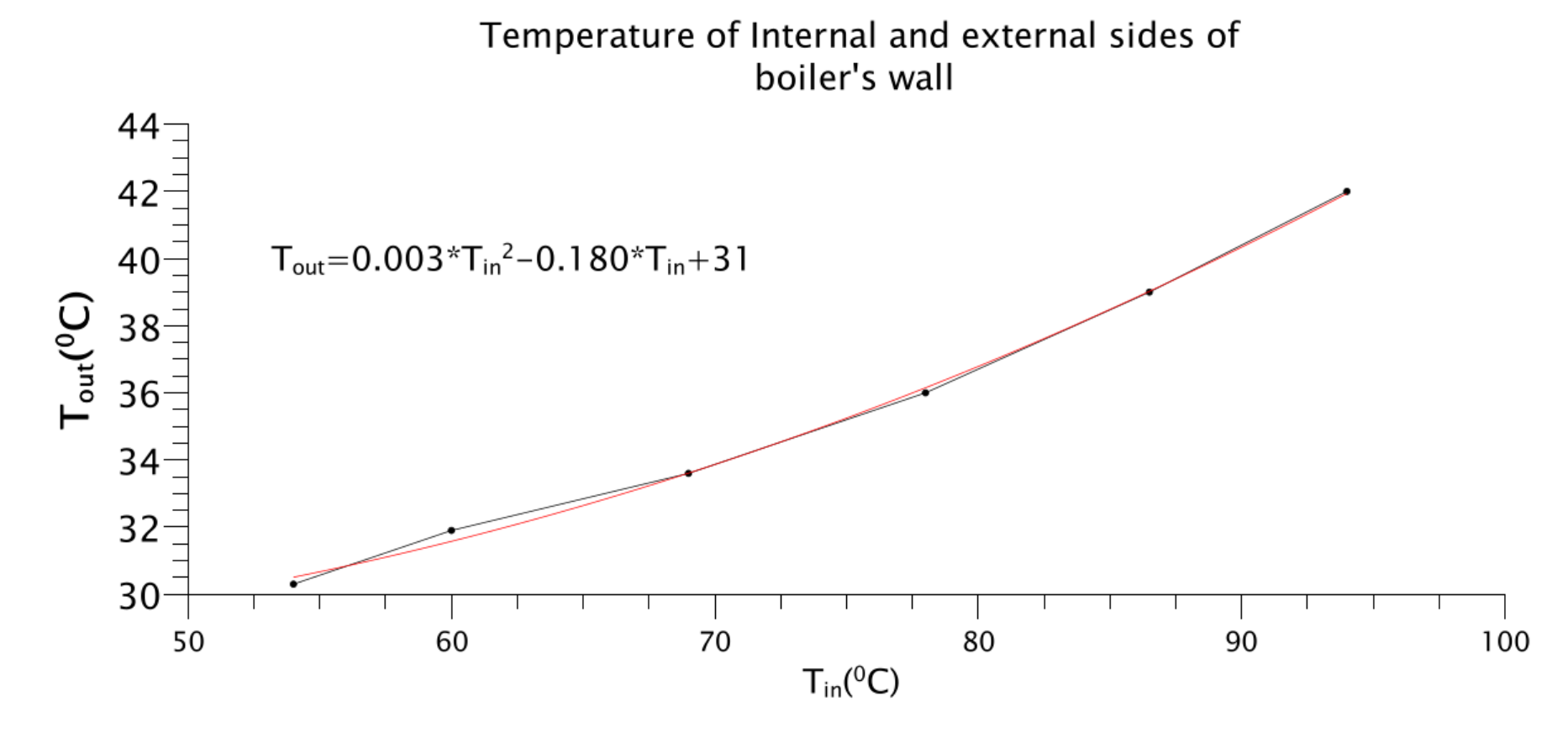}
\caption{The boiler steel wall temperature response of both sides of the cooling process. }
\end{figure}


In contrast, Figure 8 illustrates the response of aluminum, where it becomes evident that the nonlinearity in temperature response is even more pronounced. However, the magnetic response in aluminum is significantly improved compared to steel.

\begin{figure}[ht]
\centering
\includegraphics[width=\linewidth,height=8.87cm]{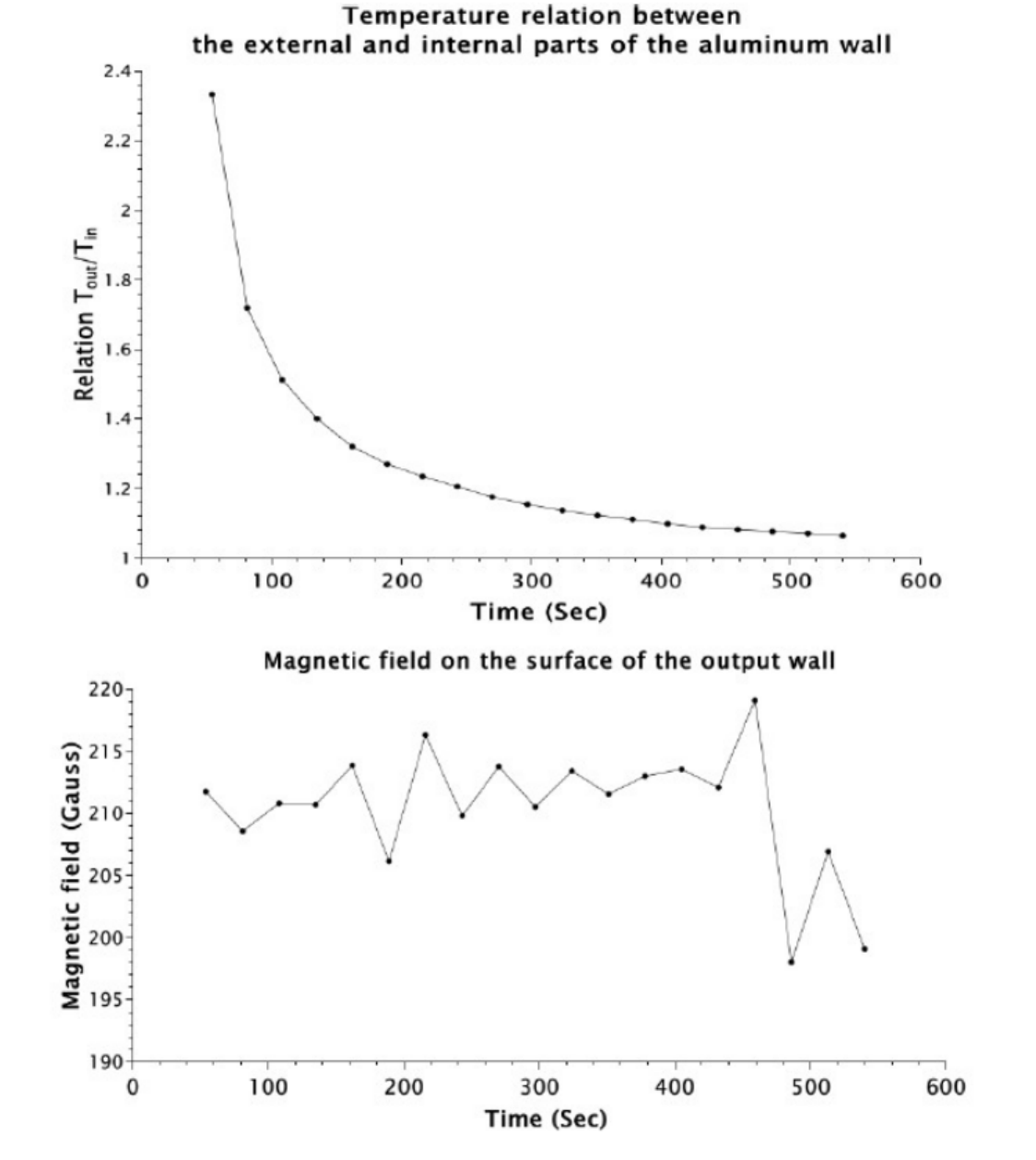}
\caption{The aluminum wall temperature response of both sides of the cooling process.}
\end{figure}

\vspace{1\baselineskip}

The results clearly show that we can use the noninvasive temperature sensor we developed in the case of nonmagnetic metal walls in the container. As we see, this method can efficiently measure inside water boilers. However, there are several important considerations to address.
Firstly, we need to assess the bimetal strip's durability in the face of corrosion, which is a potential concern. Additionally, the oxidation of one of the metals within the strip could alter its magnetic properties. Protecting the bimetal strip with an anti-oxidant coating is crucial to mitigate this. Furthermore, it's essential to investigate the durability of this bimetal strip under repeated temperature fluctuations. Finally, we must evaluate how additional materials, such as the enamel cover and thermal insulation, might affect the sensor's performance. These considerations are vital for ensuring the reliability and longevity of this noninvasive temperature sensing method.

\section{Conclusion}
Our newly developed temperature sensor is suitable for measuring the temperature of fluids within nonmagnetic containers. If corrosion challenges are overcome, our detector can be quickly implemented in water boilers and will be a solution for intelligent water temperature management.

\section*{Acknowledgments}
The authors thank Prof. Yuri Gorodetski for fruitful discussions.

\bibliographystyle{unsrt}
\bibliography{references}

\end{document}